\documentclass[10pt,twocolumn,letterpaper]{article}

\usepackage[pagenumbers]{cvpr} 

\usepackage[numbers,sort,compress]{natbib}
\usepackage{graphicx}
\usepackage{amsmath}
\usepackage{amssymb}
\usepackage{booktabs}
\usepackage[table]{xcolor}
\usepackage{arydshln}

\usepackage[pagebackref,breaklinks,colorlinks]{hyperref}
\usepackage[accsupp]{axessibility} 

\usepackage[capitalize]{cleveref}
\crefname{section}{Sec.}{Secs.}
\Crefname{section}{Section}{Sections}
\Crefname{table}{Table}{Tables}
\crefname{table}{Tab.}{Tabs.}

\newcommand{\comment}[1]{}
\newcommand{\prjwebsite}{\url{https://yorkucvil.github.io/Noise2NoiseFlow/}}

\newcommand{\ploss}{\mathcal{L}}
\newcommand{\dnloss}{\mathcal{L}_{dn}}
\newcommand{\nmloss}{\mathcal{L}_{nm}}
\newcommand{\denoise}{\mathbf{D}}
\newcommand{\noisemodel}{\mathbf{m}}
\newcommand{\img}{\mathbf{I}}
\newcommand{\cleanimg}{\img}
\newcommand{\estcleanimg}{\hat{\img}}
\newcommand{\noisyimg}{\tilde{\img}}
\newcommand{\noise}{\mathbf{N}}

\newcommand{\dnparams}{\theta}
\newcommand{\nmparams}{\phi}

\newcommand{\pnoise}{p_{\noisyimg}}

\def\cvprPaperID{6635} 
\def\confName{CVPR}
\def\confYear{2022}

\begin{document}

\title{Noise2NoiseFlow: Realistic Camera Noise Modeling without Clean Images}

\author{Ali Maleky\textsuperscript{1,3,\thanks{Work performed while interns at the Samsung AI Center--Toronto.}}, Shayan Kousha\textsuperscript{1,3,$^*$}, Michael S. Brown\textsuperscript{3}, Marcus A. Brubaker\textsuperscript{1,2,3}\\
\begin{tabular}{c c c}
    \textsuperscript{1}York University  & \textsuperscript{2}Vector Institute &
    \textsuperscript{3}Samsung AI Center--Toronto
\end{tabular}\\
}
\maketitle


\begin{abstract}
   Image noise modeling is a long-standing problem with many applications in computer vision. Early attempts that propose simple models, such as signal-independent additive white Gaussian noise or the heteroscedastic Gaussian noise model (a.k.a., camera noise level function) are not sufficient to learn the complex behavior of the camera sensor noise. Recently, more complex learning-based models have been proposed that yield better results in noise synthesis and downstream tasks, such as denoising. However, their dependence on supervised data (\ie, paired clean images) is a limiting factor given the challenges in producing ground-truth images. This paper proposes a framework for training a noise model and a denoiser simultaneously while relying only on pairs of noisy images rather than noisy/clean paired image data. We apply this framework to the training of the Noise Flow architecture. The noise synthesis and density estimation results show that our framework outperforms previous signal-processing-based noise models and is on par with its supervised counterpart. The trained denoiser is also shown to significantly improve upon both supervised and weakly supervised baseline denoising approaches. The results indicate that the joint training of a denoiser and a noise model yields significant improvements in the denoiser. 
\end{abstract}

\section{Introduction}

\begin{figure}
  \centering
    \includegraphics[width=0.475\textwidth]{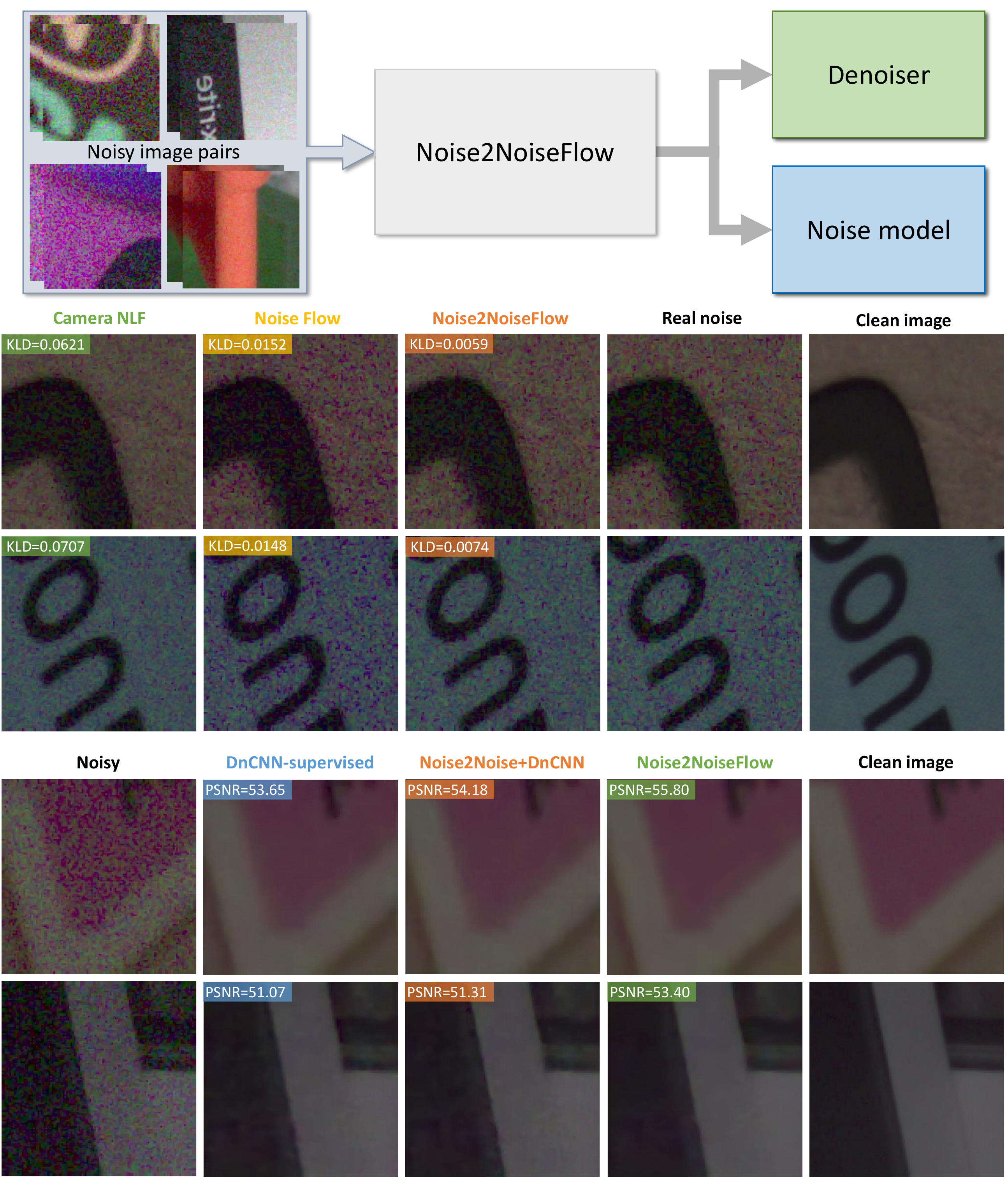}
    \caption{Overview of Noise2NoiseFlow.  (top) Given pairs of noisy images of the same scene, Noise2NoiseFlow simultaneouly trains both a noise model and a denoiser. (middle) Noise synthesis results from Camera NLF, Noise Flow, and Noise2NoiseFlow compared to the real noise in the SIDD dataset. Noise generated by Noise2NoiseFlow is the most similar to the real noise both visually and in KL divergence but without requiring clean, noise-free images.  (bottom) Example denoising results from the jointly trained denoiser compared to its supervised DnCNN baseline, and a DnCNN trained with Noise2Noise loss.}
    \label{fig:teaser}
\end{figure}

Image noise modeling is a long-standing problem in computer vision that has relevance for many applications \cite{foi2009clipped, foi2008practical, liu2014practical, liu2007automatic, chang2000adaptive, portilla2003image}.
Recently, data-driven noise models based on deep learning have been proposed \cite{henz2020synthesizing, abdelhamed2019noise, liu2021disentangling}.
Unfortunately, these models generally require clean (\ie, noise-free) images, which are practically challenging to collect in real scenarios \cite{abdelhamed2018high}.
In this work we propose a new approach, Noise2NoiseFlow, which can accurately learn noise models without the need for clean images.
Instead, only pairs of noisy images of a fixed scene are required.

While efforts are made to reduce noise during capture, post-capture modeling is a critical piece of many downstream tasks and in many domains large amounts of noise are intrinsic to the problem—for example, astro-photography and medical imaging.
As a result, noise is an integral and significant part of signal capture in many imaging domains, and modeling it accurately is critical.
For instance, noise model estimation is necessary for removing fixed pattern effects from CMOS sensors \cite{healey1994radiometric} and enhancing video in extreme low-light conditions \cite{wang2019enhancing}.
Noise models can also be used to train downstream tasks to be robust in the presence of realistic input noise.
Most naturally, they can also be used to train noise reduction algorithms without the need to collect pairs of clean and noisy images \cite{abdelhamed2019noise, nam2016holistic, zhang2021rethinking}.
However, as mentioned in \cite{zhou2020awgn, seybold2013towards, anaya2018renoir} denoisers trained with unrealistic noise models—for example, simple Gaussian noise—may not perform well on real data.

Early attempts at noise modeling were limited and failed to fully capture the characteristics of real noise.
Simple IID Gaussian noise (also called a homoscedastic Gaussian noise) ignores the fact that photon noise is signal-dependent.
Heteroscedastic Gaussian noise (\eg, \cite{foi2009clipped}) captures this by modeling noise variance as a linear function of clean image intensity but does not take into account the spatial non-uniformity of noise power, amplification noise, quantization effects, and more.
More recently, Noise Flow \cite{abdelhamed2019noise} was proposed as a new parametric structure that uses conditional normalizing flows to model noise in the camera imaging pipeline.
This model is a combination of unconditional and conditional transformations that map simple Gaussian noise into a more complex, signal-, camera-, and ISO-dependent noise distribution and outperformed previous baselines by a large margin in the normalizing flows \cite{kobyzev2021review} framework.
However, it required supervised noise data—namely, pairs of clean and noisy images—in order to learn the noise model.
Unfortunately gathering supervised data consisting of corresponding clean and noisy images can be challenging \cite{abdelhamed2018high, plotz2017benchmarking, anaya2018renoir, xu2018real} and is a limiting factor in the realistic characterization of noise.
This is even worse for other downstream tasks, which typically require large amounts of data for training.


In the context of image denoising specifically, there has been significant recent interest in methods that avoid the need for supervised data, either from careful collection or synthesis.
The well-known BM3D method~\cite{dabov2007image} proposed a denoising scheme based on transform domain representation without clean image correspondence. However, the similar patch search step makes the inference time complexity inefficient for large-scale datasets.
Recently, \citet{lehtinen2018noise2noise} introduced the Noise2Noise framework, which allowed for training of a denoiser given pairs of noisy images of the same underlying image signal.
Following this work, several others were proposed aiming to further reduce the data requirements; in particular Noise2Void \cite{krull2019noise2void} and Noise2Self \cite{batson2019noise2self} allow training of a denoiser with only individual noisy images by forcing the denoiser to predict the intensity of each pixel using only its neighbours.
Other methods attempted to add additional noise to noisy input images  \cite{pang2021recorrupted, moran2020noisier2noise, xu2020noisy} or use unpaired images in a GAN framework \cite{cha2019gan2gan, chen2018image, hong2020end, jang2021c2n, kim2019grdn}.
However, in all cases these methods are aimed primarily at denoising instead of noise modeling.

In this work, we aim to leverage these recent advances in training denoisers without direct supervision in the context of noise modeling.
Specifically, we extend the Noise2Noise framework to train a noise model with pairs of independently sampled noisy images rather than clean data.
The resulting approach, called Noise2NoiseFlow and illustrated in Figure \ref{fig:teaser}, produces both a denoiser and an explicit noise model, both of which are competitive with or out-perform fully supervised training of either model individually.

\section{Background}

Image noise can be described as an undesirable corruption added to an underlying clean signal.
Formally,
\begin{equation}
    \noisyimg = \cleanimg + \noise,
\end{equation}
where $\cleanimg$ is the underlying and mostly unobserved clean image and $\noise$ is the unwanted noise corrupting the signal, and their addition results in the noisy observation $\noisyimg$.
Different noise models are then defined by the choice of distribution assumed for $\noise$.
A widely used noise model assumes that $\noise(x,y) \sim \mathcal{N}(0,\sigma^2)$—namely, that the noise at each pixel is drawn from a zero-mean Gaussian distribution with some fixed variance.
This model has commonly been used to train and test denoisers; however, it fails to capture significant aspects of real noise, most prominently the signal-dependent variance, which is a result of the inherent Poisson shot noise \cite{liu2014practical, mohsen1975noise}.   
A significant improvement over this is heteroscedastic Gaussian noise (HGN) \cite{foi2008practical, foi2009clipped, liu2014practical} which assumes that the variance of the noise at each pixel is a linear function of the clean image intensity.
That is $\noise(x,y) \sim \mathcal{N}(0,\sigma^2(\cleanimg(x,y)))$, where $\sigma^2(\cleanimg) = \beta_1\cleanimg + \beta_2$ and $\beta_1, \beta_2$ are parameters.
This model is also sometimes referred to as the ``noise level function'' (NLF).
Recent work has shown that NLF parameters from camera manufacturers are often poorly calibrated \cite{zhang2021rethinking}; however, the NLF neglects important noise characteristics, including spatial correlation, defective pixels, clipping, quantization, and more.

To address the limitations of these pixel-independent, Gaussian-based noise models, \citet{abdelhamed2019noise} proposed the Noise Flow model, a parametric noise model based on conditional normalizing flows specifically designed to capture different noise components in a camera imaging pipeline.
In particular, Noise Flow can be seen as a strict generalization of HGN due to its use of a signal-dependent transformation layer.
However, unlike HGN, Noise Flow is capable of capturing non-Gaussian distributions and complex spatial correlations.

More recently, the DeFlow model \cite{wolf2021deflow} was proposed to handle a broader range of image degradations beyond traditional noise.
Other approaches consider mixture models or Generative Adversarial Networks (GAN) to simulate noisy and clean images in the context of  denoiser training \cite{cha2019gan2gan, chen2018image, zhu2016noise, hong2020end, jang2021c2n, kim2019grdn, henz2020synthesizing}.
However, these models are typically focused on denoising as opposed to noise modeling.
Further, GANs do not have tractable likelihoods, making the quality of the synthesized noise difficult to assess.
Most importantly, the above methods require clean images, and potentially pairs of noisy and corresponding clean images for training.
In this work we construct a formulation that explicitly trains a noise model without the need for clean images.
Because of the flexibility and generality of the normalizing flow framework and quality of its results, we will focus on the Noise Flow model \cite{abdelhamed2019noise} here, though, as we will discuss, other choices are possible.

\subsection{Image Denoising}

Image noise reduction has been a long-standing topic of study in computer vision \cite{kuan1985adaptive, liu2007automatic, zhang2008multiresolution, chang2000adaptive, portilla2003image, dabov2007image}.
Here we focus on recent methods that have found success by leveraging large training sets and deep learning architectures \cite{zhang2017beyond}.
These methods are characterized by regressing, typically with a convolutional neural network, from a noisy image observation to its clean counterpart.
Given a training set $\mathcal{D} = \{(\noisyimg^{(i)}, \cleanimg^{(i)})\}_{i=1}^N$ of noisy images $\noisyimg$ and their corresponding clean images $\cleanimg$, learning of a denoiser $\denoise$ is then formulated as minimizing
\begin{equation}
    \sum_{i=1}^N\ploss(\denoise(\noisyimg^{(i)};\dnparams), \cleanimg^{(i)}),
\end{equation}
where $\ploss$ is typically an $L_1$ or $L_2$ norm and $\denoise$ is a deep neural network with parameters $\dnparams$.

This approach is limited by the need to have access to the corresponding clean image $\cleanimg$, and several notable approaches have recently been explored to remove this requirement.
Most relevant to this work is the Noise2Noise framework, proposed by \citet{lehtinen2018noise2noise}.
Rather than requiring clean/noisy pairs of images, it simply requires two noisy observations of the same underlying clean signal.
Given a dataset of noisy image pairs $\{(\noisyimg_1^{(i)}, \noisyimg_2^{(i)})\}_{i=1}^N$, the Noise2Noise framework optimizes the loss function
\begin{equation}
    \sum_{i=1}^N\ploss(\denoise(\noisyimg_1^{(i)};\dnparams), \noisyimg_2^{(i)}) + \ploss(\denoise(\noisyimg_2^{(i)};\dnparams), \noisyimg_1^{(i)})\ .
    \label{eq:n2n_loss}
\end{equation}
That is, the the second noisy image is used as the target for the denoiser of the first and vice versa.
Perhaps surprisingly, training with this objective is still able to produce high-quality denoising results, despite the lack of access to clean images  \cite{lehtinen2018noise2noise}.
In this work, we aim to explore the generalization of this approach to noise modeling.

\section{Noise2NoiseFlow}

\begin{figure*}[!t]
    \includegraphics[width=\textwidth]{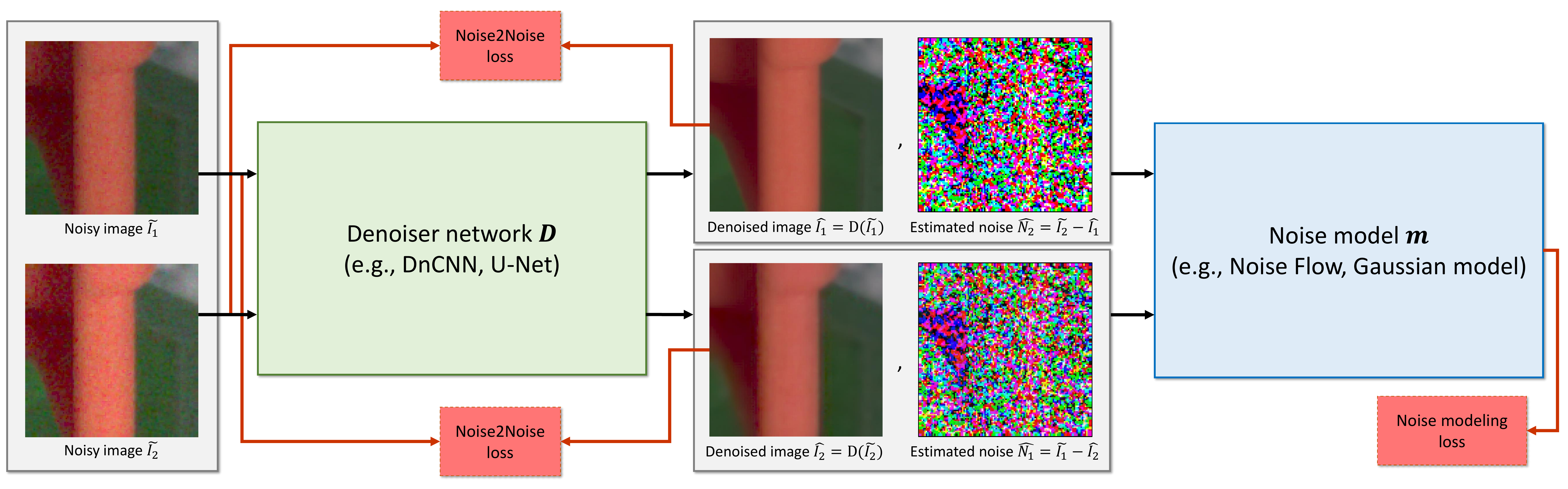}
    \caption{An overview of the training loss for the proposed Noise2NoiseFlow framework. Given a pair of independent noisy samples of the same underlying signal, it runs both noisy samples through a denoiser network $\denoise$, which outputs the estimated clean signal.  We then use the estimated clean image from the first image in place of the true clean signal for the second noisy observation and vice versa.  This prevents the denoiser from collapsing into a degenerate solution of an identity transformation.  Note that in the paper we formulate the noise model as a distribution over the noisy image $\noisyimg$, though, as shown here, it is common for noise models to be expressed as a distribution of the residual noise $\noise = \noisyimg - \cleanimg$.  These two formulations are equivalent.
}
    \label{fig:n2nf_overview}
\end{figure*}

\comment{{{
\begin{figure*}[!t]
    \includegraphics[width=\textwidth]{figures/overview - Itr 8.pdf}
    \caption{The overview of our proposed generic Noise2NoiseFlow framework. It takes a pair of independent noisy samples of the same scene $(\noisyimg_1,\noisyimg_2)$ and runs it through a denoiser network $\denoise$, that outputs estimates of the clean signals corresponding to each of the noisy input images. The result is two training pairs $(\noisyimg_1, \denoise(\noisyimg_2;\dnparams)$ and $(\noisyimg_2, \denoise(\noisyimg_1;\dnparams)$ which is used to calculate $\pnoise(\noisyimg_1 | \denoise(\noisyimg_2;\dnparams) ; \nmparams)$ and $\pnoise(\noisyimg_2 | \denoise(\noisyimg_1;\dnparams) ; \nmparams)$ using the noise model $\noisemodel$. $\dnparams$ and $\nmparams$ correspond to the parameters of $\denoise$ and $\noisemodel$.}
    \label{fig:n2nf_overview}
\end{figure*}
}}}

In this section, we define our approach to learning a noise model with weak supervision—namely, through the use of only pairs of noisy images.
There are two main components, a denoiser $\denoise(\cdot ; \dnparams)$, which learns to predict the clean image $\cleanimg$ given a noisy image, $\noisyimg$, as input, and a model of a noisy image given the clean image $\cleanimg$, $\pnoise(\cdot | \cleanimg ; \nmparams)$.
The denoiser and noise model have parameters $\dnparams$ and $\nmparams$ respectively.
Our goal is to learn the distribution $\pnoise(\noisyimg | \cleanimg)$—namely, the distribution of noisy image conditioned on the clean image—without explicitly requiring $\cleanimg$.\footnote{Note that this is equivalent to learning the distribution of the noise conditioned on the clean image by simply shifting the distribution of noise by the clean image.}
To do this, we propose to use the output of the denoiser as an estimate of the clean image—That is, $\cleanimg \approx \estcleanimg = \denoise(\noisyimg ; \dnparams)$.
We could in principle then learn $\pnoise$ by minimizing $-\log \pnoise ( \noisyimg | \estcleanimg ; \nmparams)$ with respect to the noise model parameters $\nmparams$.
However, this requires a well-trained denoiser, which, in turn, typically requires access to clean images to train.
Further, if we tried to simultaneously train the denoiser and noise model, there is a trivial singular optimum where the denoiser converges to the identity and the noise model converges to a Dirac delta at zero.

Drawing inspiration from the Noise2Noise framework \cite{lehtinen2018noise2noise}, we instead assume we have access to pairs of noisy observations $\noisyimg_1,\noisyimg_2$ which both have the same underlying clean signal, $\cleanimg$.
That is, $\noisyimg_1 = \cleanimg + \noise_1$ and $\noisyimg_2^{(i)} = \cleanimg + \noise_2$, where $\noise_1$ and $\noise_2$ are independent samples of noise.
Then, given the pairs of noisy images, we can use the denoiser applied to one image to estimate the clean image for the other image in the pair.
That is, we propose to optimize the loss

{\footnotesize 
\begin{equation}
    \nmloss(\noisyimg_1,\noisyimg_2) = - \log \pnoise(\noisyimg_1 | \denoise(\noisyimg_2;\dnparams) ; \nmparams) - \log \pnoise(\noisyimg_2 | \denoise(\noisyimg_1;\dnparams) ; \nmparams)
    \label{eq:nm_loss}
\end{equation}
}
for both the noise model parameters $\nmparams$ and the denoiser parameters $\dnparams$.
Because the two images are of the same underlying scene, the output of the denoiser should ideally be the same for both noisy images.
However, because the two images have independent samples of noise, the denoiser cannot simply collapse to the identity.
This is analogous to the Noise2Noise objective, where the output of the denoiser on one image is used as the target for the other image in the pair.
In practice, we find it beneficial to include the Noise2Noise objective function to stabilize the training of the denoiser together with the noise model objective.
That is, we propose to train the denoiser and noise model jointly with the loss $\ploss(\noisyimg_1,\noisyimg_2) = \nmloss(\noisyimg_1,\noisyimg_2) + \lambda \dnloss(\noisyimg_1,\noisyimg_2)$, where
\begin{equation}
    \dnloss(\noisyimg_1,\noisyimg_2) = \lVert \denoise(\noisyimg_1;\dnparams) - \noisyimg_2 \rVert_2^2 + \lVert \denoise(\noisyimg_2;\dnparams) - \noisyimg_1 \rVert_2^2
    \label{eq:dn_loss}
\end{equation}
is the Noise2Noise loss.
Given a dataset of pairs of noisy images, $\mathcal{D} = \{(\noisyimg_1^{(i)}, \noisyimg_2^{(i)})\}_{i=1}^N$, we optimize the loss over the set of pairs 
\begin{equation*}
\sum_{i=1}^N \ploss(\noisyimg_1^{(i)},\noisyimg_2^{(i)}),
\end{equation*}
where the optimization can be done with a stochastic optimizer.
In this work we use Adam \cite{kingma2014adam}.

Figure \ref{fig:n2nf_overview} shows an overview of the proposed approach.
We note that the formulation is generic to the choice of denoiser and noise model, requiring only that the noise model's density function can be evaluated and that both the noise model and denoiser can be differentiated as needed.
In the experiments that follow we primarily use the DnCNN architecture \cite{zhang2017beyond} for the denoiser, as it is a standard denoiser architecture based on residual connections and convolutional layers.
For the noise model we primarily focus on Noise Flow \cite{abdelhamed2019noise} due to its flexibility and tractability and, consequently, dub our proposed method \textbf{Noise2NoiseFlow}.
However, we also explore other choices for these components, such as a U-Net architecture for the denoiser and the heteroscedastic Gaussian noise model.

\section{Experiments}
Here we explore the performance of the proposed Noise2NoiseFlow approach.
To do this we make use of Smartphone Image Denoising Dataset (SIDD) \cite{abdelhamed2018high} to assess the accuracy of both our learned noise model and the image denoiser.
SIDD contains images of 10 different scenes consisting of a range of objects and lighting conditions, which were captured with five different smartphone cameras at a range of different ISO levels.
Multiple captures of each scene instance were taken and carefully aligned in order to produce a corresponding ``clean'' image for each noisy image.
While our proposed method does not require the clean images for training, we do make use of them for a quantitative evaluation against a range of baselines, including methods that require clean image supervision.
Here we use two different subsets of SIDD—namely SIDD-Full and SIDD-Medium.
While SIDD provides both sRGB and rawRGB images, here we only consider the rawRGB images. 
SIDD-Full provides 150 different noisy captures for each corresponding clean image.
In contrast, SIDD-Medium contains only a single noisy image for each clean image. 
To extract the noisy/noisy image pairs of the same clean signal from SIDD-Full that are required by our method for training, we select pairs of noisy images corresponding to the same clean image.
In order to maximize alignment between the selected two images, we select consecutive images from the 150 available for each scene in SIDD-Full.

We use SIDD-Medium to evaluate the performance of our method.
Specifically, while we use noisy/noisy pairs of images extracted from SIDD-Full for training as described above, we evaluate the performance of both the denoiser $\denoise(\cdot)$ and the noise model $\pnoise(\cdot|\cleanimg)$ using the noisy/clean image pairs in SIDD-Medium.
In order to test Noise2NoiseFlow against our baselines, we use supervised noisy/clean pairs from SIDD-Medium.
Denoting $(\noisyimg, \cleanimg)$ as a noisy/clean image pair, we evaluate the noise modeling using the negative log-likelihood per dimension $D^{-1} \log \pnoise(\noisyimg  | \cleanimg; \nmparams)$, where $D$ is the total number of dimensions (both pixels and channels) in the input.
Negative log likelihood is a standard evaluation metric for generative models and density estimation, but it is known to be less sensitive to distributions that overestimate the variance of a distribution. 
To account for this we also evaluate the model using the Kullback-Leibler (KL) divergence metric introduced in \cite{abdelhamed2019noise}.
Both NLL and KL divergence are reported in nats.
Specifically, given a noisy and clean image, we compute a histogram of both real noise and noise generated by a model by subtracting the clean image and computing the KL divergence between the two histograms.
See \cite{abdelhamed2019noise} for more details on this metric.
To evaluate the denoiser, we compute peak signal-to-noise ratio (PSNR) and the structural similarity index measure (SSIM).

SIDD contains scenes with ISO levels ranging from 50 to 10,000; however, many of those ISO levels have only a small number of images available.
To be consistent with other methods that use SIDD for noise modeling—for example, \cite{abdelhamed2019noise}—we remove images with rare ISO levels, keeping only ISO levels 100, 400, 800, 1600, and 3200.
After filtering, approximately 500,000 patches of size 32$\times$32 pixels are extracted.
The extracted patches are separated into training and test sets using the same training and testing split of SIDD scenes that was used in \cite{abdelhamed2019noise}.
Approximately 70\% of the extracted patches were used for training and the remaining were used as testing.
We trained all models using the Adam optimizer \cite{kingma2014adam} for 2,000 epochs.
We used a value of $\lambda = 2^{18}$ in all experiments, unless otherwise noted.
To speed up convergence and avoid early training instabilities we pre-trained the denoiser $\denoise$ on the training set using $\dnloss$ alone for all of the experiments.
The architecture of the Noise Flow noise model and DnCNN denoiser was the same as in \cite{abdelhamed2019noise}, but both were reimplemented in PyTorch and verified to produce equivalent results as the original Noise Flow implementation.

\subsection{Noise Modeling}
We first compare our proposed approach quantitatively to traditional noise models which have been calibrated using supervised, clean images.
Table \ref{tab:synthesis_modeling} compares the results of our model against the camera noise level function (Cam-NLF), a simple additive white Gaussian noise model (AWGN), and Noise Flow \cite{abdelhamed2019noise}.
Despite only having access to pairs of noisy images, the proposed Noise2NoiseFlow has effectively identical performance to the state-of-the-art Noise Flow model which is trained on clean/noisy image pairs.
To demonstrate the benefit of joint training, we trained a Noise2Noise denoiser \cite{lehtinen2018noise2noise} on noisy/noisy paired data and use this to denoise images to train Noise Flow.  We refer to this as ``N2N+NF.''

We also compared our results to the recently released ``calibrated Poisson-Gaussian'' noise model described in \cite{zhang2021rethinking}.
The results for this comparison in terms of KL divergence can be found in Table \ref{tab:per_camera} for the three cameras reported in the paper \cite{zhang2021rethinking}, as the Calibrated P-G model included noise parameters only for three different sensors: iPhone 7, Samsung Galaxy S6 Edge, and Google Pixel.
It is clear that while the Calibrated P-G model improves over the in-camera noise level function, it still lags behind both Noise Flow and Noise2NoiseFlow.
We again see that the proposed Noise2NoiseFlow outperforms this very recent method.

\begin{figure*}[!t]
\centering
\includegraphics[width=0.93\textwidth]{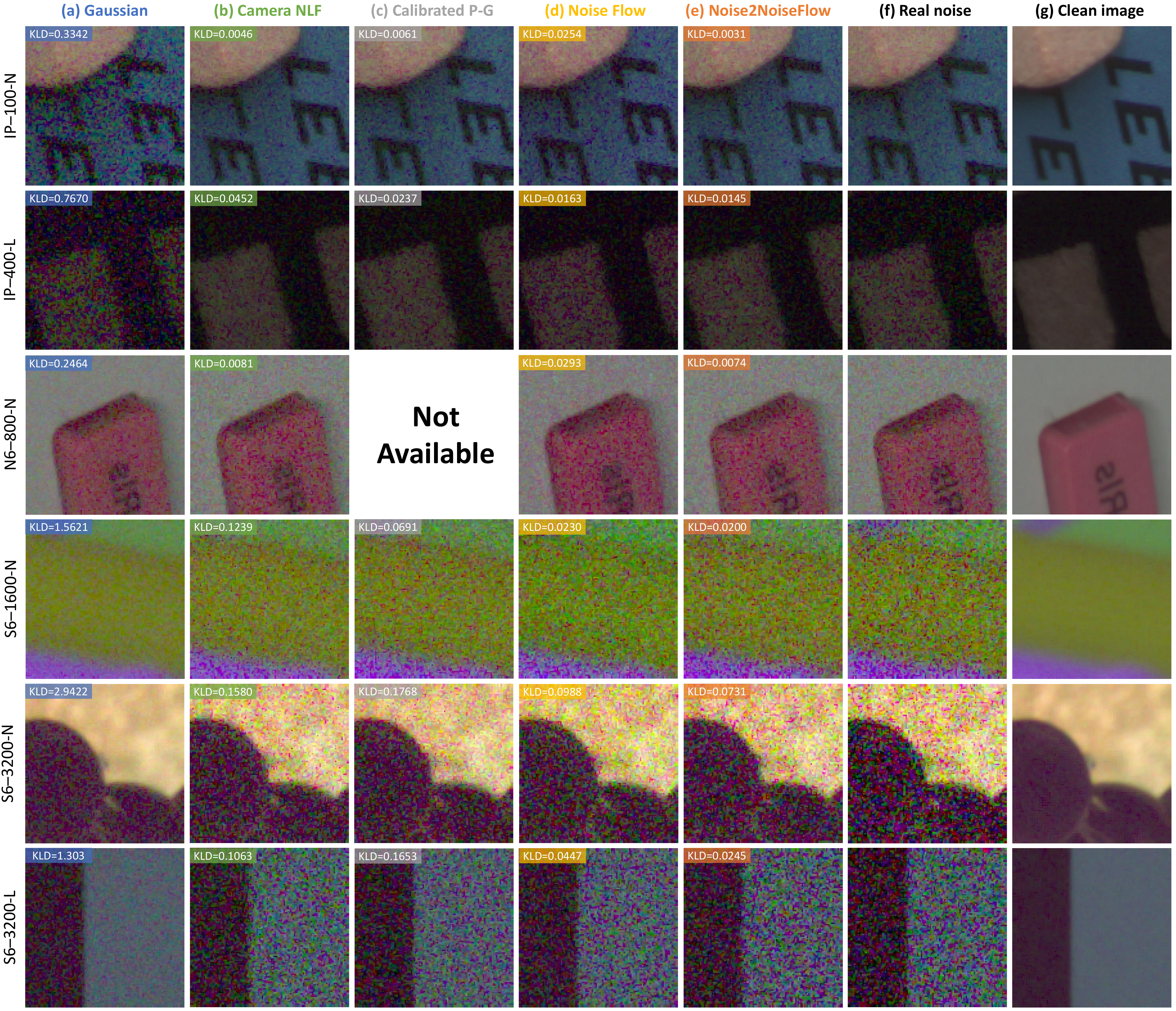}
\caption{Noise synthesis samples from (a) the AWGN model, (b) Camera NLF, (c) Calibrated P-G \cite{zhang2021rethinking}, (d) Noise Flow \cite{abdelhamed2019noise}, and our proposed method, Noise2NoiseFlow, compared to the (f) real noise in SIDD. Our samples are closer to real noise in terms of both visual perception and the KL divergence metric. Codes on the left indicate [camera]-[ISO]-[brightness]. For Calibrated P-G \cite{zhang2021rethinking} calibrated parameters for SIDD only include three camera sensors (IP, GP, and S6) so some synthesis results are not available.}
\label{fig:synthesis}
\end{figure*}

Figure \ref{fig:synthesis} shows qualitative noise samples generated by Noise2NoiseFlow, as well as other baselines compared to the real noise. The samples are generated for different camera sensors, ISO levels, and scenes.
The suffix N corresponds to normal light and L corresponds to the low-light conditions.
As evidenced by these images, the results from Noise2NoiseFlow are both visually and quantitatively better than other baselines, especially in low-light/high-ISO settings, where other baselines underperform.

\begin{table}[!t]
  \centering
  \begin{tabular}{ c c c }
 \hline
 Model & $NLL$ & $D_{KL}$ \\
   \hline
   AWGN & -2.874 &
 0.4815 \\
 Cam. NLF & -3.282 &
 0.0578 \\
 N2N+NF & -3.459 &
 0.0363 \\
 Noise Flow & -3.502 &
 0.0267 \\
 \hdashline
 Noise2NoiseFlow & 
 -3.501 & 0.0265 \\
 \hline
\end{tabular}
\caption{Negative log-likelihood per dimension and $D_{KL}$ results on test data for baseline models and our proposed Noise2NoiseFlow model. Noise2NoiseFlow significantly improves over AWGN and Camera NLF and is on par with the Noise Flow model, while requiring no clean images.
It also improves over separately training a Noise2Noise denoiser and NoiseFlow (\it{N2N+NF}), demonstrating the value of joint training.
  \label{tab:synthesis_modeling}}
\end{table}

\begin{table}[!t]
  \centering
  \begin{tabular}{c c c c | c}
 \hline
  & IP & S6 & GP & Agg \\
   \hline
   AWGN & 0.4353 &
 0.4863 & 0.5865 & 0.4934\\
 Cam. NLF & 0.0513 &
 0.1014 & 0.0212 & 0.0596\\
 Calibrated P-G & 0.0188 &
 0.0981 & 0.0332 & 0.0492\\
 Noise Flow & 0.0112 &
 0.0469 & 0.0180 & 0.0250\\
 \hdashline
 Noise2NoiseFlow & 
 0.0125 & 0.0444 & 0.0190 & 0.0249\\
 \hline
\end{tabular}
\caption{Per camera KL divergence performance of our model Noise2NoiseFlow compared to the baselines on three camera sensors for which the Calibrated P-G model is defined as well as the aggregate results on the test data for these three sensors.}
  \label{tab:per_camera}
\end{table}

\subsection{Noise Reduction}
\begin{figure}[!t]
\centering
\includegraphics[width=0.478\textwidth]{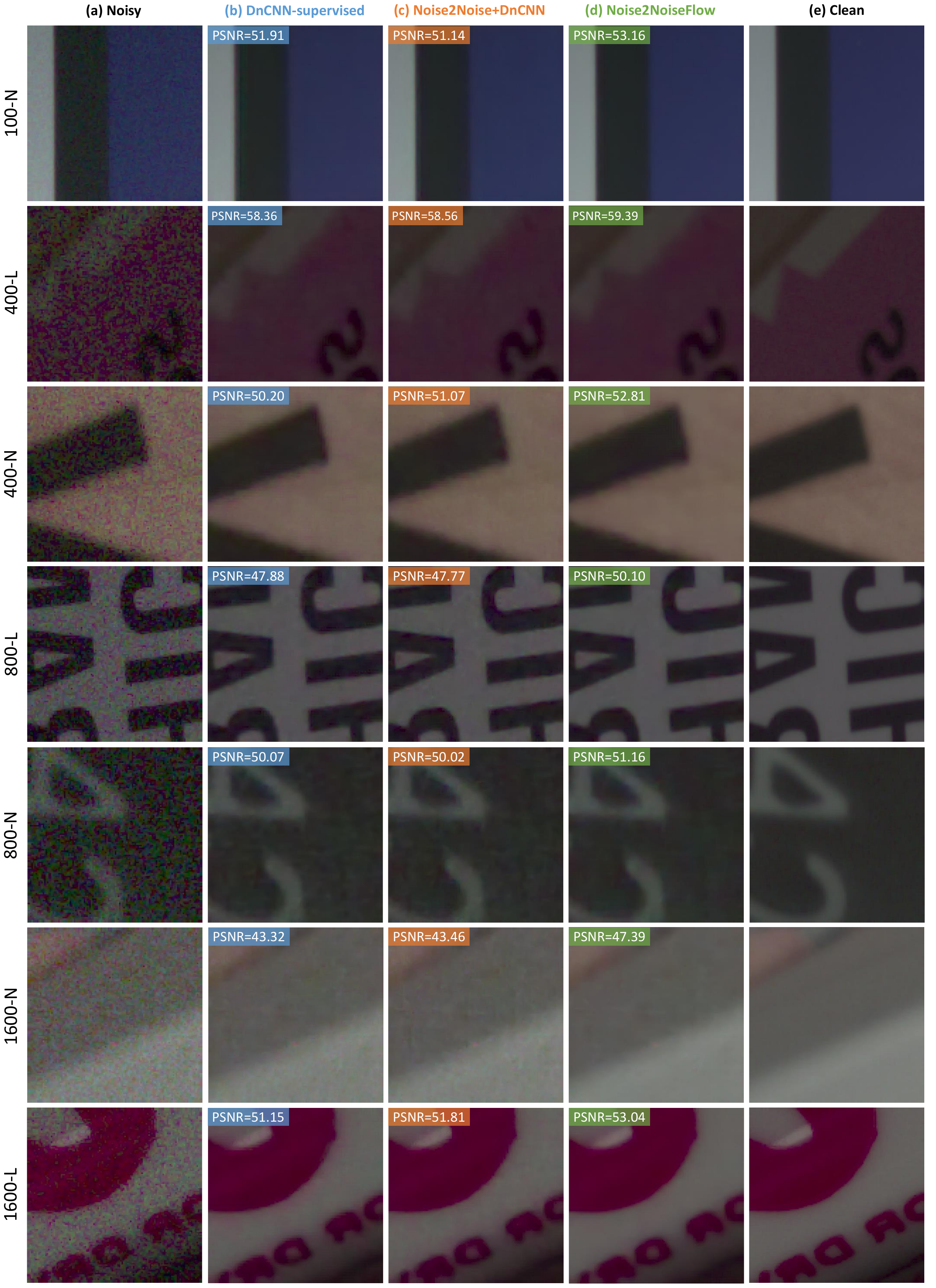}
\caption{Denoising results from (b) DnCNN-supervised, (c) Noise2Noise trained with DnCNN, and (d) Noise2NoiseFlow on samples from SIDD-Medium testing data. The codes on the left indicate the ISO level as well as the lighting condition.}
\label{fig:denoising}
\end{figure}

While the primary goal of this work was noise modeling, it also includes a denoiser as a key component.
Here we investigate the performance of the denoiser by evaluating its performance in terms of PSNR on the held-out test set.
We compared against three scenarios, which are reported in Table \ref{tab:denoising}.
In all cases the exact same DnCNN architecture is used.
First, we trained the same denoiser architecture $\denoise$ using the Noise2Noise \cite{lehtinen2018noise2noise} loss alone.
This is shown in Table \ref{fig:denoising} as ``Noise2Noise+DnCNN'' and shows that, indeed, the joint noise model training improves the denoising performance by over 1.2dB, a significant margin in PSNR.
Second, we trained a supervised DnCNN model using the corresponding clean image patches for the training set; this is indicated in the table as ``DnCNN-supervised''.
Noise2NoiseFlow outperforms this by nearly 1.5dB, despite not having access to clean images.
In fact, both Noise2Noise+DnCNN and Noise2NoiseFlow outperform this clean-image supervised baseline, suggesting that the increased variety of data available with noisy image pairs appears to be more valuable than access to clean images.
We also trained a supervised Noise Flow model and used samples generated from the model to train a DnCNN denoiser. We refer to this baseline as ``DnCNN - NF synthesized''. The ``DnCNN - NF synthesized'' outperforms the ``DnCNN-supervised'' baseline which is consistent with the results reported in the Noise Flow paper \cite{abdelhamed2019noise}.  However, it still significantly underperforms Noise2NoiseFlow.

Figure \ref{fig:denoising} shows qualitative denoising results from Noise2NoiseFlow and the aforementioned baselines.
The results show that our model performs better in denoising, especially in more severe situations (high ISO and low brightness).
The estimated clean signal tends to be much smoother and cleaner for Noise2NoiseFlow than both of its baselines in terms of visual perception and PSNR in almost all the cases.
Taken together, our results demonstrate that the joint training of both an explicit noise model and a denoiser not only allows for weakly supervised training, but also improves the resulting estimated denoiser.

\begin{table}[!t]
  \centering
  \begin{tabular}{ c c c }
 \hline
 Model & PSNR & SSIM \\
   \hline
 Noise2Noise+DnCNN & 51.57 & 0.977 \\
 DnCNN-supervised & 51.32 & 0.980 \\
 DnCNN - NF synthesized & 51.71 & 0.980 \\
 \hdashline
 Noise2NoiseFlow(ours) & \textbf{52.80} & \textbf{0.984}\\
 \hline
\end{tabular}
\caption{Denoising results from a DnCNN trained with supervised noisy/clean paired data from SIDD-Medium, Noise2Noise with a DnCNN trained on nosiy/noisy image pairs, a DnCNN trained with noise samples generated from a supervised Noise Flow model, and our denoiser model trained on the same noisy/noisy data measured by PSNR and SSIM on the test set.}
  \label{tab:denoising}
\end{table}

\subsection{Ablation Studies}
\begin{figure}
  \centering
    \includegraphics[width=0.475\textwidth]{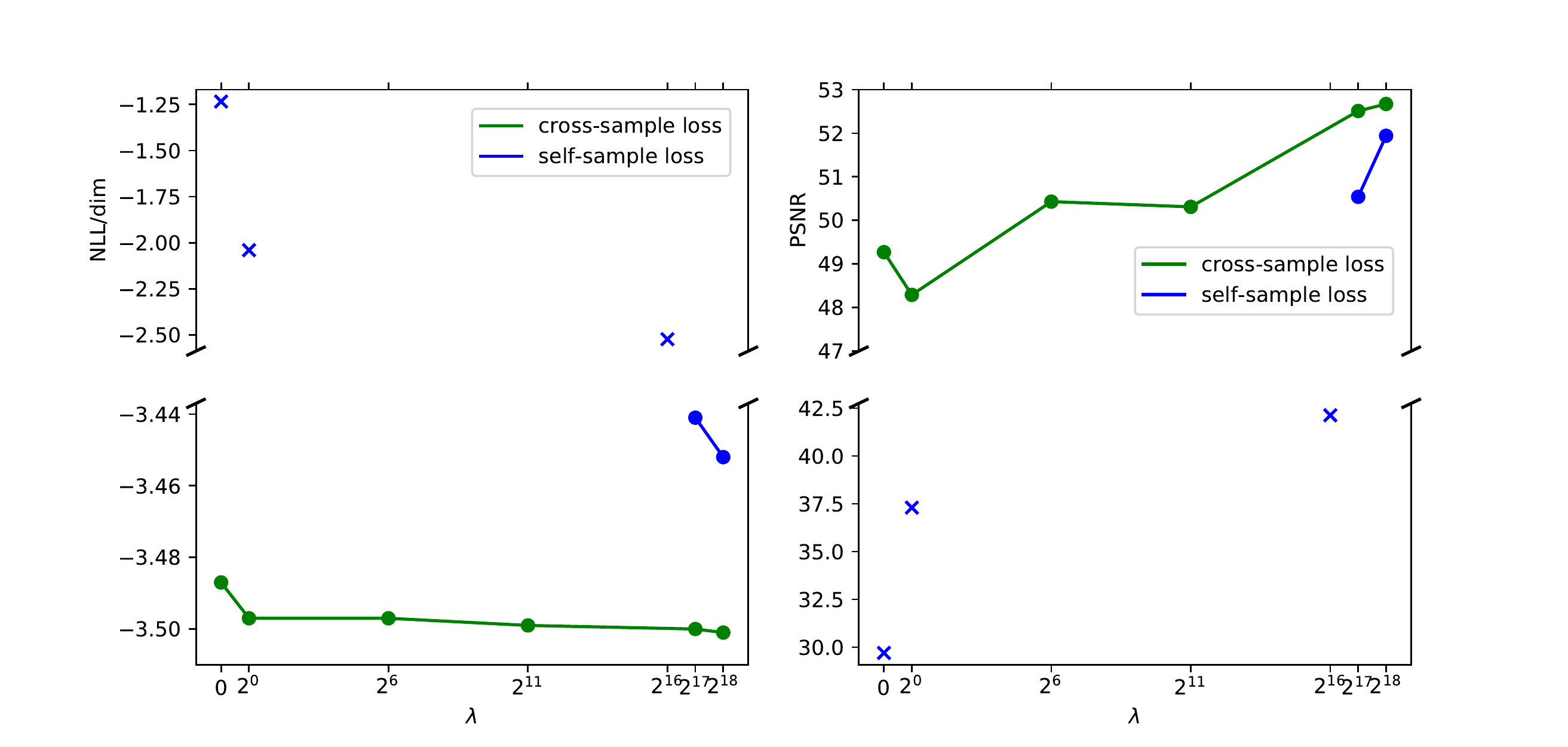}
    \caption{Negative log-likelihood per dimension and PSNR results on test data as a function of the regularization term $\lambda$. Cross-sample loss refers to training with Eq.~\ref{eq:nm_loss} and self-sample loss refers to training with Eq.~\ref{eq:xnm_loss}. Cross marks indicate the loss at the last epoch in cases where training failed as evidenced by spikes in NLL and significant drops in PSNR. All experiments with the self-sample loss and $\lambda \leq 2^{16}$ ultimately diverge.}
    \label{fig:loss_fn_ablation}
\end{figure}

We next investigate the design choices for our framework and their impact on the results.
First, we conduct an ablation on the value of $\lambda$, the weighting factor for the Noise2Noise loss.
We explored a wide range of values, from $\lambda=0$ to $\lambda=2^{18}$.
For each value, we computed the negative log-likelihood per dimension and the PSNR of the denoiser.
The results are plotted in Fig.~\ref{fig:loss_fn_ablation} and show that our results are relatively robust to the choice of $\lambda$.
While a value of $\lambda=0$ produces reasonable results, better results are generally obtained with larger values of $\lambda$.
This indicates that the Noise2Noise loss in Eq.~\ref{eq:dn_loss} plays an important role in stabilizing the training and ensuring consistency of the denoiser.

Next, we consider a different form of the loss function where we use the estimated clean image based on $\noisyimg_1$ for the noise model loss function with $\noisyimg_1$.
Formally, we use the noise model objective
{\footnotesize 
\begin{equation}
\nmloss(\noisyimg_1,\noisyimg_2) =
- \log \pnoise(\noisyimg_1 | \denoise(\noisyimg_1;\dnparams) ; \nmparams)
- \log \pnoise(\noisyimg_2 | \denoise(\noisyimg_2;\dnparams) ; \nmparams)
\label{eq:xnm_loss}
\end{equation}
}
instead of the one proposed in Equation \ref{eq:nm_loss}.
We refer to training based on this model as the self-sample loss, in comparison to the cross-sample loss.
While a seemingly innocuous change, training based on Equation \ref{eq:xnm_loss} becomes extremely unstable.
In this case, the denoiser can converge to a degenerate solution of the identity function—namely, $\denoise(\noisyimg) = \noisyimg$—which allows the noise model $\pnoise$ to converge to a Dirac delta and the value of $\nmloss(\noisyimg_1,\noisyimg_2)$ goes to negative infinity.
This behaviour can be alleviated with large values of $\lambda$, which can be seen in Figure \ref{fig:loss_fn_ablation}, where settings of $\lambda$ that resulted in diverged training are indicated with a cross at the value of the epoch before the divergence occurred.
As the figure shows, values $\lambda$ less than $2^{17}$ resulted in this behaviour.
In contrast, the proposed loss function in Equation \ref{eq:nm_loss} is robust to the choice of $\lambda$, even allowing training with a value of $\lambda=0$, which disables the $\dnloss$ term from Equation \ref{eq:dn_loss} entirely.
We also explored higher values for $\lambda$ (\eg, $2^{19}$) but did not observe significant changes in behaviour.

We also explored different choices of denoiser architecture and noise model as our framework is agnostic to these specific choices.
For the denoiser, beyond the DnCNN architecture, we also considered the U-Net \cite{ronneberger2015u} denoiser architecture used in \cite{lehtinen2018noise2noise}.
For the noise model, beyond the Noise Flow-based model, we also considered the heteroscedastic Gaussian noise model, or noise level function (NLF), due to its ubiquity.
We implemented the NLF as a variation on a Noise Flow architecture.
Specifically, taking the signal-dependent and gain layers of the Noise Flow model, without any of the other flow layers, results in a model that is equivalent to the NLF.

The results of this experiment can be found in Table \ref{tab:nf_nlf}, which reports the negative log likelihood per dimension, KL divergence metric, and PSNR of the resulting noise model and denoiser for all combinations of these choices.
The results indicate that the choice of denoiser architecture is not particularly important.
Both U-Net and DnCNN produce similar results to one another, for both choices of noise model.
However, we see that the use of the Noise Flow model over the heteroscedastic Gaussian noise model does provide a boost in performance for both noise modeling and denoising.
Further, and consistent with results reported recently elsewhere \cite{zhang2021rethinking}, we see that a retrained heteroscedastic Gaussian noise model can outperform the parameters provided by camera manufacturers.

\begin{table}[t]
  \centering
  \begin{tabular}{ c c | c c c }
 \hline
 Denoiser & Noise model & $NLL$ & $D_{KL}$ & PSNR\\
   \hline
 DnCNN & Noise Flow & -3.501 &
 0.0265 & 52.80 \\
 U-Net & Noise Flow & 
 -3.500 & 0.0255 & 52.64 \\
 DnCNN & NLF & 
 -3.461 & 0.0288 & 52.69 \\
 U-Net & NLF & 
 -3.463 & 0.0332 & 52.13 \\
 \hline
\end{tabular}
\caption{Performance on test data from our ablated models. Each row corresponds to a specific choice for the noise model and the denoiser architecture. NLF corresponds to our modified version of heteroscedastic Gaussian model implemented as a bijective normalizing flow transformation.}
  \label{tab:nf_nlf}
\end{table}

\subsection{Training with Individual Noisy Images}

Here we have proposed a novel approach to noise model training by coupling the training of a noise model with a denoiser and based on the Noise2Noise framework.
This naturally raises the question of whether a noise model could be trained with only individual noisy images, particularly given the success of such approaches for denoisers.
All of these approaches aim to prevent the denoiser from collapsing into the degenerate solution of an identity transformation, similar to the behaviour identified above with the alternative loss formulation in Equation \ref{eq:xnm_loss}, by either using a blind-spot network architecture (\eg, Noise2Void \cite{krull2019noise2void} and Noise2Self \cite{batson2019noise2self}), or adding additional noise to the input images (\eg, Noisier2Noise \cite{moran2020noisier2noise}, Noisy-as-Clean \cite{xu2020noisy}, and R2R \cite{pang2021recorrupted}).
To investigate this idea we considered using the R2R \cite{pang2021recorrupted} framework, which, given a single noisy image $\noisyimg$, generates two new noisy images as 
\begin{equation}
    \noisyimg_{\text{input}} = \noisyimg + \mathbf{D}^{T}\mathbf{z}, \quad  \noisyimg_{\text{target}} = \noisyimg - \mathbf{D}^{-1}\mathbf{z},
    \label{eq:r2r_np}
\end{equation}
where $\mathbf{z}$ is drawn from $\mathcal{N}(0,I)$, and $\mathbf{D} = \alpha I$ is an invertible matrix with scale parameter $\alpha$.
We modify our loss functions to utilize these new images so that $\nmloss = -\log \pnoise(\noisyimg | \denoise(\noisyimg_{\text{input}}))$ and $\dnloss = \Vert \noisyimg_{\text{target}} - \denoise(\noisyimg_{\text{input}}) \Vert_2^2$ and train by optimizing $\ploss = \nmloss + \lambda \dnloss$ as described above.
We use the same DnCNN architecture for $\denoise$ and the Noise Flow model for $\pnoise$ and report the results in Table \ref{tab:r2rf}, with this variation labelled as R2RFlow and compared against a clean-image supervised Noise Flow model and the noisy-pair supervised Noise2NoiseFlow.
The results indicate that the R2RFlow approach yields a reasonable noise model, though significantly below the performance of Noise2NoiseFlow, particularly in terms of denoising.
However, the experiment is enticing and suggests that this is a promising direction for future work.

\begin{table}[t]
  \centering
  \begin{tabular}{ c c c c }
 \hline
 Model & $NLL$ & $D_{KL}$ & PSNR \\
   \hline
 Noise Flow & -3.502 &
 0.0267 & N/A \\
 Noise2NoiseFlow & 
 -3.501 & 0.0265 & 52.80 \\
 R2RFlow & 
 -3.443 & 0.0983 & 50.08 \\
 \hline
\end{tabular}
\caption{Performance on test data from our proposed R2RFlow formulation that requires only single noisy samples and no supervision in any forms compared to Noise2NoiseFlow (weak supervision) and the Noise Flow model (supervised).}
  \label{tab:r2rf}
\vspace*{-0.25cm}
\end{table}

\section{Conclusions and Future Work}

We introduced a novel framework for jointly training a noise model and denoiser that does not require clean image data.
Our experimental results showed that, even without the corresponding clean images, the noise modeling performance is largely the same when training only with pairs of noisy images.
We believe this approach can improve the practicality of the existing noise models in real-world scenarios by reducing the need to collect clean image data, which can be a challenging, tedious, and time-consuming process and may not be possible in some settings, \eg, medical imaging.
Further, joint training was shown to improve denoising performance when compared with a denoiser trained alone.
The learned denoiser can even surpass supervised baselines, which we hypothesize is due to the increased number of noisy images and indicating that noise modeling can provide useful feedback for denoising.

While training a noise model without clean image data is a significant step towards more practical noise models, our proposed approach still required paired noisy images.
We believe that it may be possible to go further still and train a noise model in a purely  unsupervised manner, \ie, without clean images or pairs of noisy images.
Our preliminary experiments with the R2R framework \cite{pang2021recorrupted} suggest that this may indeed be feasible, but more work remains to be done.
Code for this paper is available at: \prjwebsite.
{\small
\paragraph*{Acknowledgements}
This work was done during an internship at the Samsung AI Center in Toronto, Canada.  AM's internship was funded by a Mitacs Accelerate.  SK's and AM's student funding came in part from the Canada First Research Excellence Fund for the Vision: Science to Applications (VISTA) programme and an NSERC Discovery Grant. 
}



{\small
\setlength{\bibsep}{0pt plus 0.3ex}
\bibliographystyle{IEEEtranSN}
\bibliography{main}
}
\clearpage

\newpage
\appendix
\section{Training details}

In this section, we give more details about the training procedure. As mentioned in the main paper, we used Adam \cite{kingma2014adam} as optimizer in all of our experiments. We pre-trained the denoiser with N2N loss (Eq. 5 of the main paper) for 2,000 epochs. Also note that the denoiser pre-training step was used only to boost training under different setups, and is not a vital part of the overall training. Training the original Noise2NoiseFlow model from scratch will also produce almost the same results ($NLL$: $-3.498$, $D_{KL}$: $0.0275$, PSNR: $52.65$).

The supervised DnCNN was trained with MSE using the clean/noisy pairs from SIDD-Medium. Both denoiser pretraining and supervised training used an initial learning rate of $10^{-3}$, which was decayed to $10^{-4}$ at epoch 30, and $5 \times 10^{-5}$ at epoch 60. We used orthogonal weight initialization \cite{hu2020provable} for the denoiser architectures and the exact same initial weights for the noise model as used in the Noise Flow paper.

\begin{figure}
  \centering
    \includegraphics[width=0.475\textwidth]{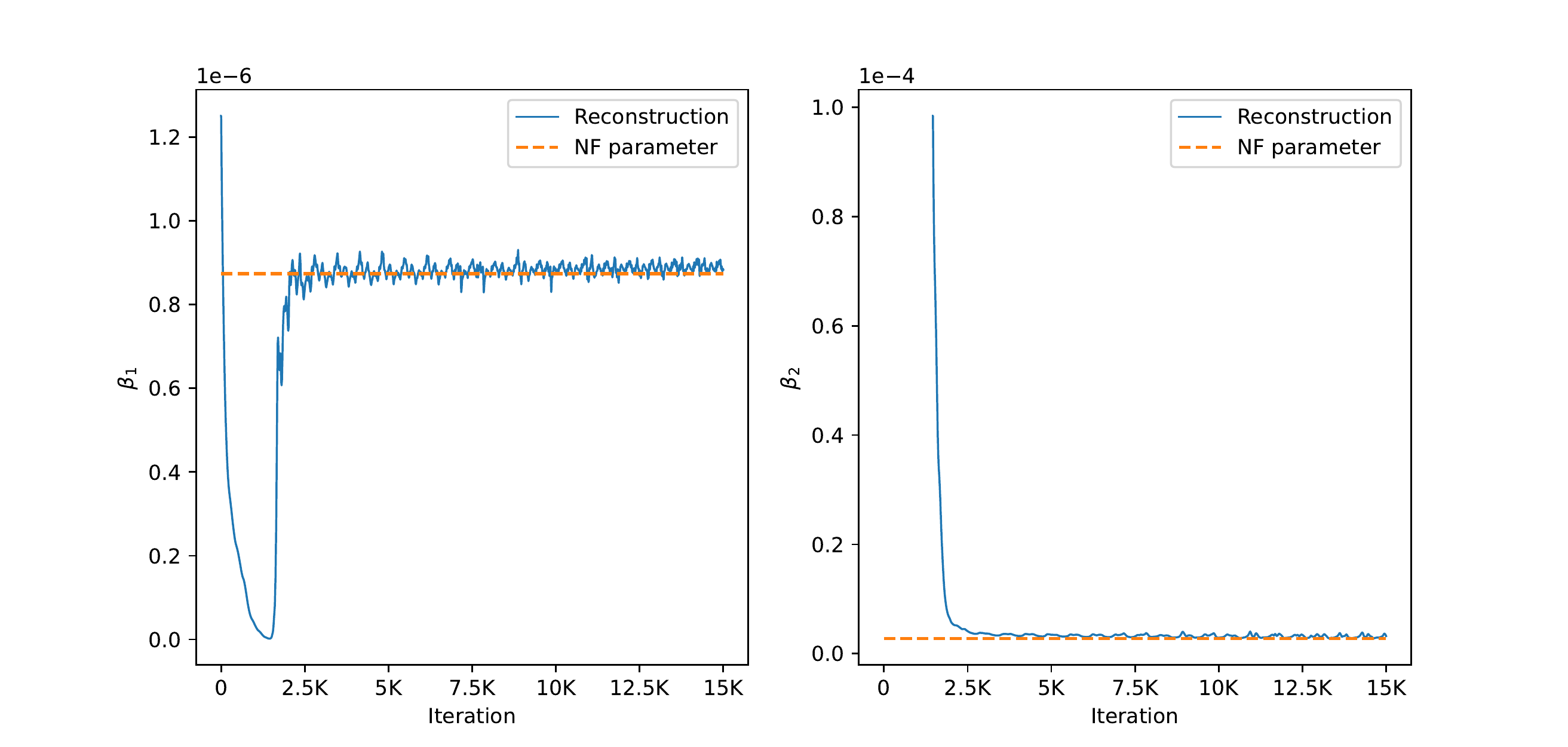}
    \caption{Convergence curve of the two parameters ($\beta_1$ and $\beta_2$ of the NLF model for a specific camera sensor and ISO level. \emph{NF Parameter} corresponds to the parameters learned by a supervised Noise Flow model and \emph{Reconstruction} Corresponds to the NLF parameters learned by a Noise2NoiseFlow model from synthetic data generated by the supervised Noise Flow model. As evidenced by the figures, the model can successfully retrieve the parameters. }
    \label{fig:synthetic_ablation}
\end{figure}

The denoiser was a 9 layer DnCNN and was the same in all experiments except where noted. Noise Flow was re-implemented in PyTorch \cite{paszke2019pytorch} and carefully tested for consistency against the original implementation.
Joint training used a constant learning rate of $10^{-4}$ for 2,000 epochs though no improvements were generally observed after $\sim600$ epochs.

\section{Synthetic Noise Experiment}

In order to demonstrate that our framework can retrieve the parameters of a supervised trained noise model, we have conducted a synthetic noise experiment. In this setting, we first trained a heteroscedastic Gaussian noise model, which was implemented as a flow layer in Noise Flow. For simplicity, we only took one camera and one ISO setting—namely, iPhone 7 and 800 as ISO level as we had adequate image data for training and evaluation. Under the mentioned setting, the model only has two trainable parameters—namely, $\beta_1$ and $\beta_2$. We then use this trained model to synthesize noisy image pairs for training a subsequent Noise2NoiseFlow model from scratch with only a heteroscedastic Gaussian layer as its noise model and DnCNN as its denoiser. The results shown in Figure \ref{fig:synthetic_ablation} shows that our model can successfully retrieve the parameters of a trained NLF model.

\section{Failure Cases}

\begin{figure*}[!t]
\includegraphics[width=\textwidth]{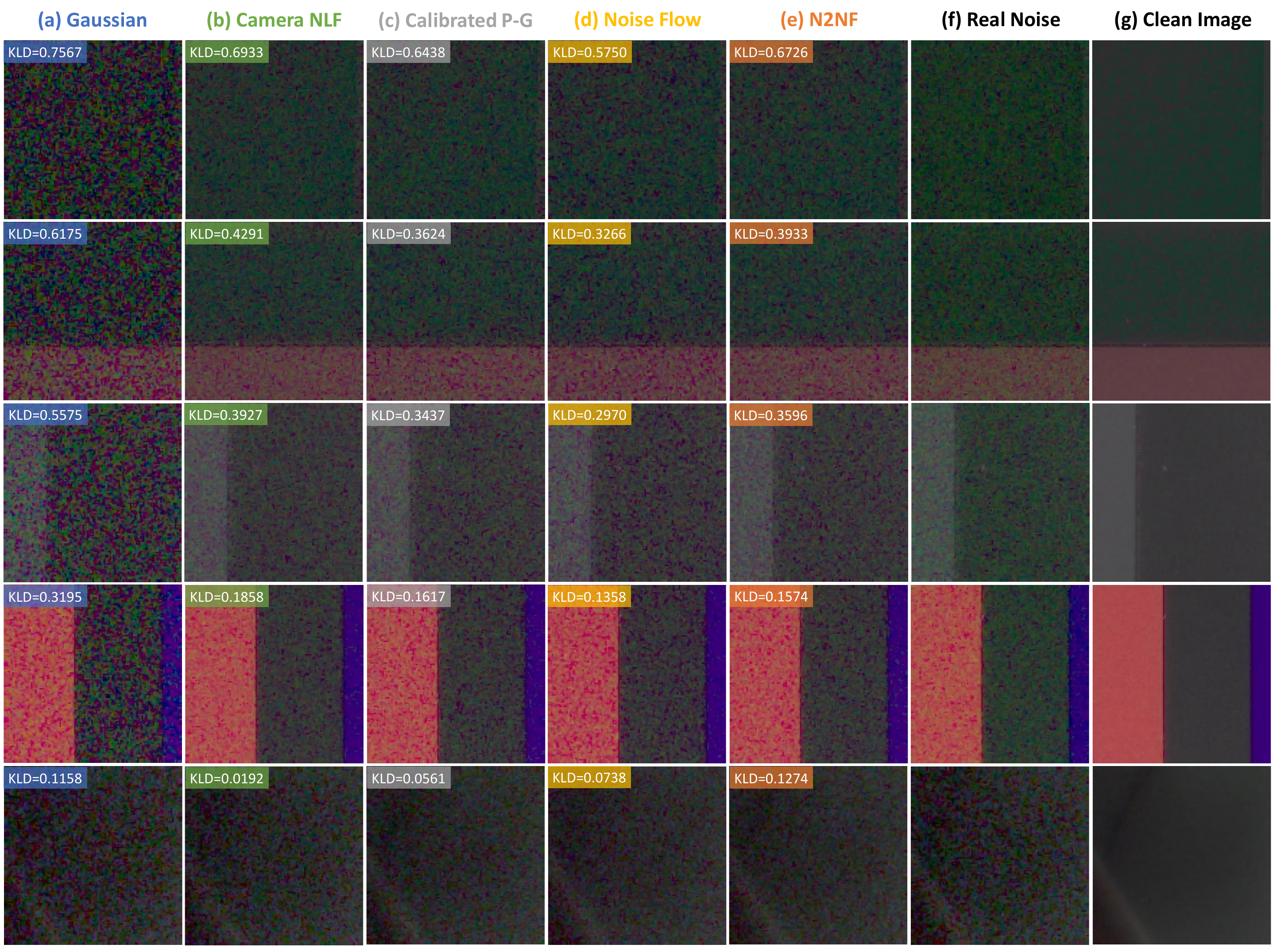}
\caption{Noise synthesis samples from (a) the AWGN model, (b) Camera NLF, (c) Calibrated P-G \cite{zhang2021rethinking}, (d) Noise Flow \cite{abdelhamed2019noise}, and our proposed method, Noise2NoiseFlow, compared to the (f) real noise in SIDD for patches where Noise2NoiseFlow has the worst $D_{KL}$ numbers.}
\label{fig:worst_samples}
\end{figure*}

Although no significant unrealistic behaviour was noticed, we visualize 5 noise samples with the worst $D_{KL}$ for Noise2NoiseFlow in Figure \ref{fig:worst_samples}. While the noise samples are not in the best alignment with the real samples, the generated noise patches do not look very unnatural.

\end{document}